\documentclass[
aps,
prl,
twocolumn,
superscriptaddress,
nofootinbib,
longbibliography
]{revtex4-2}

\usepackage{amsmath}
\usepackage{amssymb}
\usepackage{bm}
\usepackage{graphicx}
\usepackage{dcolumn}
\usepackage{booktabs}
\usepackage{xcolor}
\usepackage{hyperref}

\def\nue{\mathrel{{\nu_e}}}

\def\barnue{\mathrel{{\bar \nu}_e}}

\newcommand{\be}{\begin{equation}}
\newcommand{\ee}{\end{equation}}
\newcommand{\ba}{\begin{eqnarray}}
\newcommand{\ea}{\end{eqnarray}}

\newcommand{\n}{neutrino}
\newcommand{\ns}{neutrinos}
\newcommand{\sn}{supernova}
\newcommand{\sne}{supernovae}
\newcommand{\df}{DSNB}

\begin{document}

\title{A Two-Parameter Framework for the Diffuse Supernova Neutrino Background}

\author{Cecilia Lunardini}
\affiliation{Department of Physics, Arizona State University, Tempe, Arizona 85287, USA}

\author{Xingyun Yang}
\affiliation{Department of Physics, Arizona State University, Tempe, Arizona 85287, USA}

\begin{abstract}
Following important experimental developments, tests of the Diffuse Supernova Neutrino Background (\df) are entering a new phase that might lead to a discovery. This motivates the development of simple and broadly applicable tools for comparing theoretical predictions and interpreting experimental results. 
We show that, over the energy range relevant to current detectors, the \df\ is well described by an exponential spectrum, characterized 
by 
the differential flux at 20 MeV, $\phi_{20}$, and a spectral energy scale, $\mathcal{E}_0$. We fit the spectra of 24 representative models and demonstrate that the exponential description provides an excellent approximation. We then recast the recent Super-Kamiokande results in the $(\phi_{20},\mathcal{E}_0)$ plane, showing that model-specific best fits and upper bounds can be interpreted naturally within this framework. Our results indicate that, above realistic detection thresholds, and until the precision phase is reached, the observable \df\ is effectively a two-parameter phenomenon. We advocate that the $(\phi_{20},\mathcal{E}_0)$ parameterization be adopted as a standard, to be used routinely to report future theoretical predictions and experimental results. 
\end{abstract}

\maketitle

After the historic detection of a \n\ burst from 
SN1987A \cite{Hirata:1987hu,Bionta:1987qt,Alekseev:1988gp},
a modern observation of \ns\ from core collapse \sne\ is still a missing piece in the already rich field of \n\ astrophysics. It is possible that, in the coming years, 
this gap will be filled 
by the
observation of the Diffuse Supernova Neutrino Background (\df), the cumulative flux of \ns\ from all past stellar collapses in the Universe \cite{NYAS:NYAS319,Krauss:1984aa} (see, e.g., \cite{Suliga:2022ica} for a review). 
After decades of experimental development \cite{Malek:2003ki,Super-Kamiokande:2011lwo,SNO:2020gqd,KamLAND:2022sqb,Chernyak:2026xhc,Super-Kamiokande:2023xup}, this flux is now within the reach of current and next generation \n\ observatories. After starting its Gadolinium phase \cite{Beacom:2003nk,Koshio:2025fjs}, Super-Kamiokande (SK from here on) obtained very stringent bounds on the \df\ \cite{Super-Kamiokande:2025sxh} and, recently, it has observed an indication of it, at $2.6 \sigma$ C.L. \cite{Sekiya:Neutrino2026,SuperK:2026DSNBPressRelease}.  This update suggests that the \df\ might be discovered in the next few years, by the multi-pronged searches at Super-Kamiokande, JUNO \cite{JUNO:2022lpc}(currently running), Hyper-Kamiokande \cite{Abe:2011ts} (starting in 2027), and, in the longer term, DUNE \cite{DUNE:2020lwj}. 
The observation of the \df\ will be a milestone, marking the transition of 
\sn\ \ns\ 
to continuous data taking, and testing important phenomena like the state and evolution of the collapsed core,  neutrino flavor evolution in ultra-dense matter, and more.

A challenge for current \df\ phenomenology is the proliferation of theoretical models. Recent experimental analyses have considered dozens of benchmark spectra derived from different assumptions about stellar populations, neutrino emission, flavor conversion, and collapse outcomes. While these calculations differ in their underlying physics, many produce very similar observable spectra in the 
experimentally accessible energy window.
As a result, comparisons among models, and between models and data, are often more cumbersome than necessary. In the anticipation of the next, rich phase of \df\ studies,
it is important to develop a common, practical language that can be used by theorists, experimental collaborations, and the broader neutrino community to describe and interpret the \df.

An early attempt at such simplification was done in Ref.~\cite{Lunardini:2006pd} (after an early suggestion in \cite{Malek:2003ki}). It was shown that, if (i) the \n\ spectrum of an individual supernova is described by the (quasi-thermal) alpha spectrum \cite{Keil:2002in,Tamborra:2012ac}, and (ii) all supernovae are identical \n\ emitters, 
the \df\ spectrum above realistic detection thresholds can be approximated by an exponential form: $\phi(E) \propto e^{-E/\mathcal{E}_0}$, where $\mathcal{E}_0$  is (approximately) related to the first and second momenta of 
the single-source \n\ spectrum, $\langle E \rangle$ and $\langle E^2 \rangle$: $\mathcal{E}_0\simeq (\langle E^2 \rangle-\langle E \rangle^2)/\langle E \rangle$.

Here we develop this idea, 
and explore its connection to 
current and upcoming \df\ searches. We test if the exponential form is an adequate effective description of modern theoretical models of the \df, and discuss its applicability to current experimental results. We then 
propose it as a useful tool in 
future studies of the \df.

To fix the ideas, we consider the electron antineutrino ($\barnue$) component of the \df, which is the main target 
at water Cherenkov and liquid scintillator detectors, and restrict to the window $E=10-35$ MeV, where backgrounds are the lowest, and so detectors are most sensitive \cite{Super-Kamiokande:2025sxh,JUNO:2022lpc}. 
We adopt the parameterization 
\begin{equation}
\phi(E)
=
\phi_{20}
\exp\left[
-\frac{E-20~{\rm MeV}}{\mathcal{E}_0}
\right],
\label{eq:exp}
\end{equation}
where 
the differential flux at 20 MeV, $\phi_{20}$, was chosen because it can be directly measured, since it falls inside the detection window.

\begin{figure*}[htbp]
    \centering \includegraphics[width=0.75\linewidth]{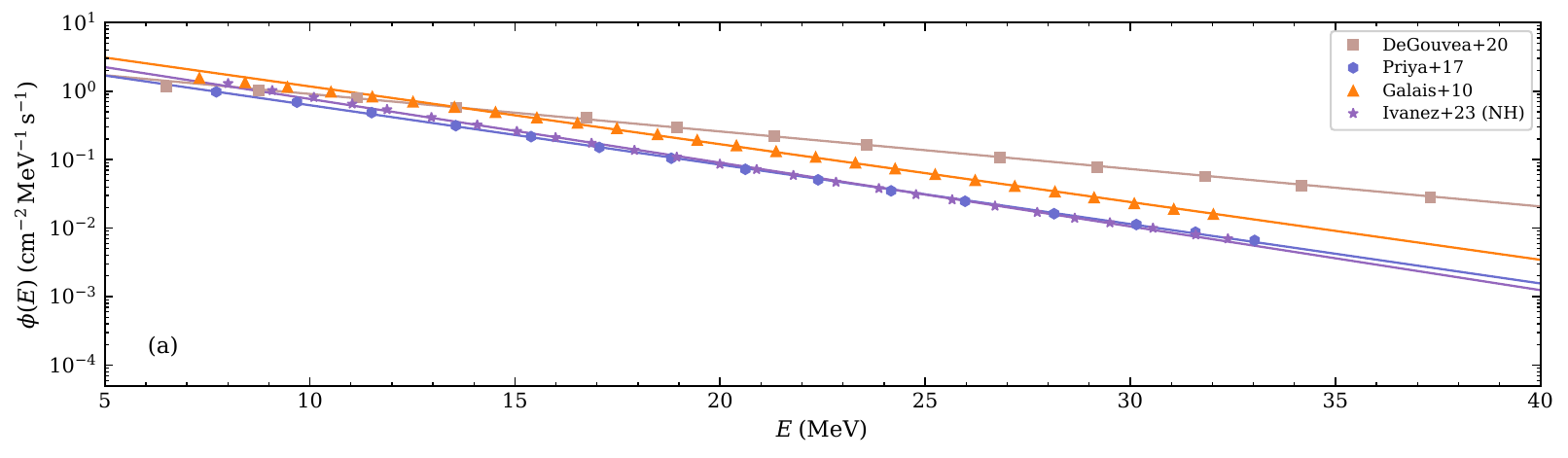}
   \vspace{-0.2cm}

    \includegraphics[width=0.75\linewidth]{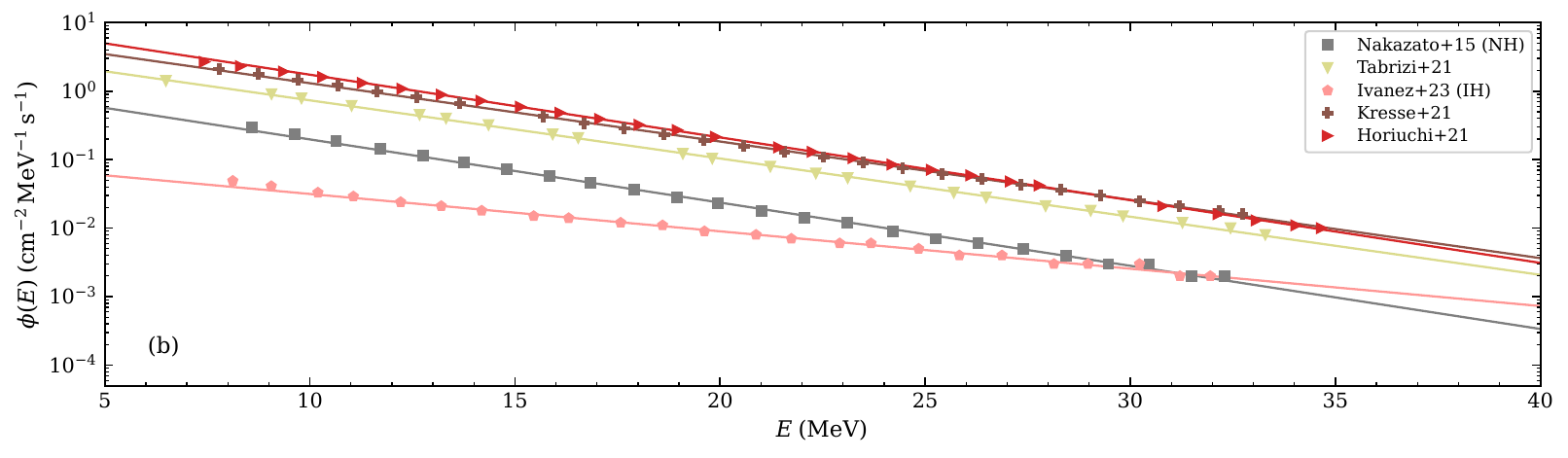}
    \vspace{-0.2cm}

    \includegraphics[width=0.75\linewidth]{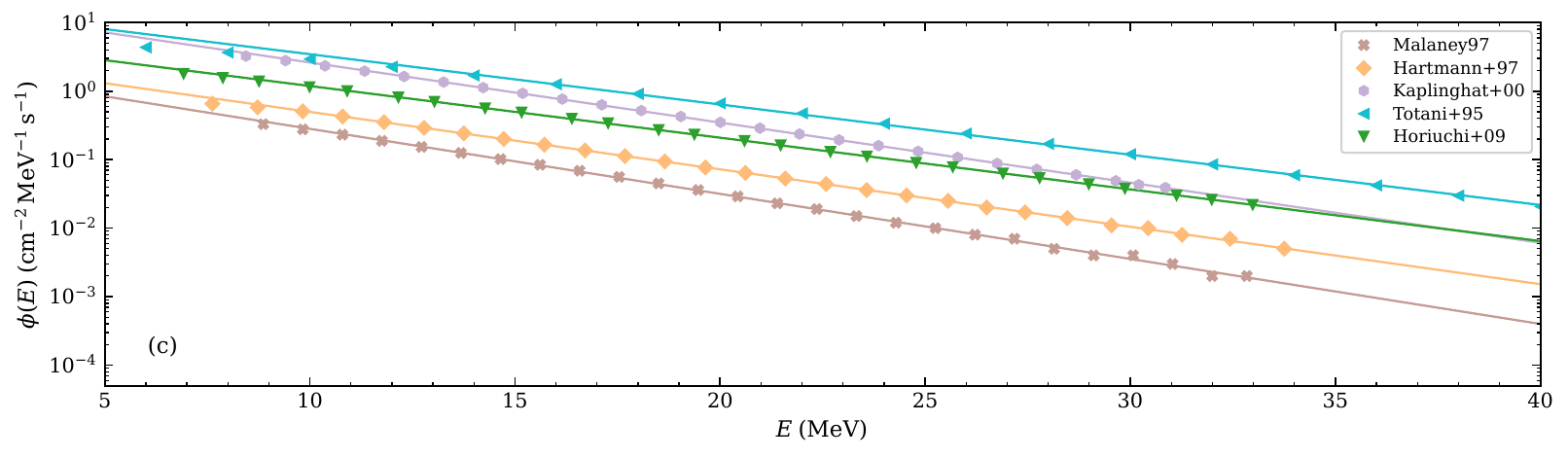}
    \vspace{-0.2cm}

    \includegraphics[width=0.75\linewidth]{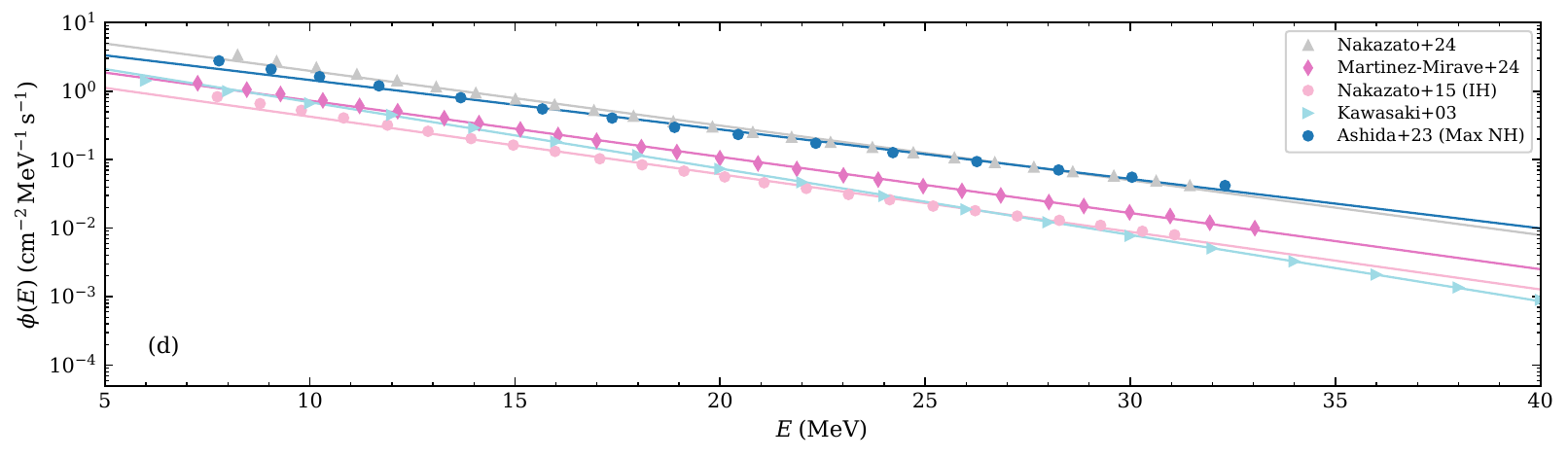}
    \vspace{-0.2cm}

    \includegraphics[width=0.75\linewidth]{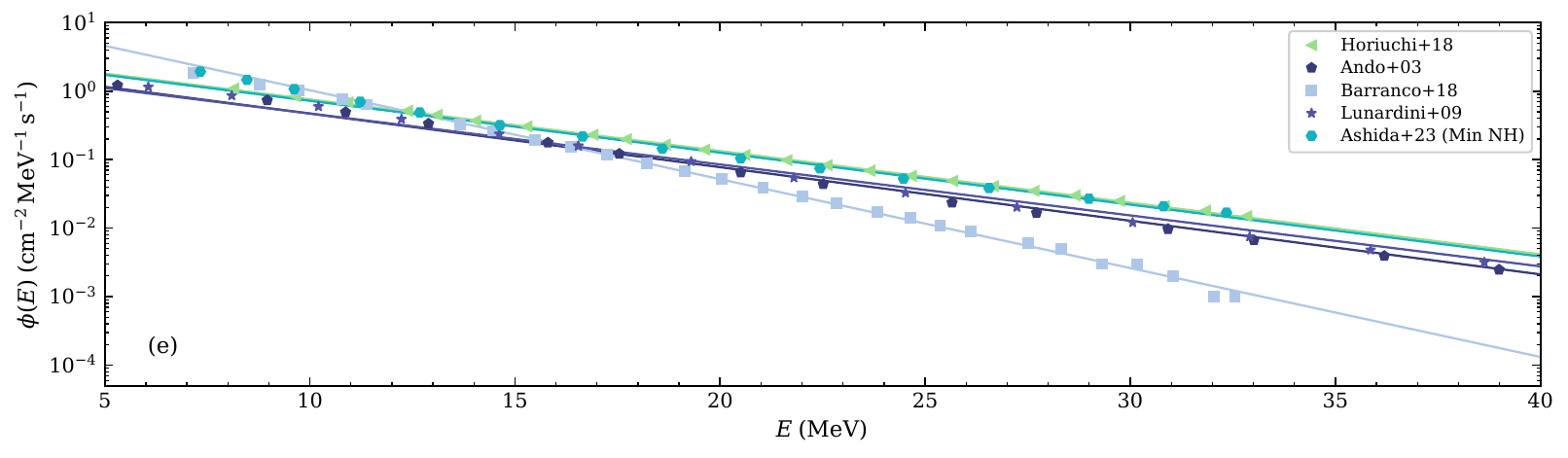}

    \caption{DSNB flux spectra and exponential fit curves for the chosen set of 24 models  
\cite{Totani:1995rg,Malaney:1996ar,Hartmann:1997qe,Kaplinghat:1999xi,Fukugita:2002qw,Ando:2005,Horiuchi:2008jz,Lunardini:2009ya,Galais:2009wi,Nakazato:2015rya,Priya:2017bmm,Horiuchi:2017qja,Barranco:2017lug,DeGouvea:2020ang,Horiuchi:2020jnc,Kresse:2020nto,Tabrizi:2020vmo,Ivanez-Ballesteros:2022szu,Ashida:2023heb,Martinez-Mirave:2024zck} (see Supplemental Material for details).   In each case, the numerically calculated points are shown together with the best-fit exponential form, Eq. (\ref{eq:exp}).
    }
    \label{fig:dsnb_expofit_five_panels}
\end{figure*}

As a first step, we
compare the template in Eq. (\ref{eq:exp}) to a large collection of \df\ models from the literature. For the sake of comparison with experimental results, we consider a set of 24 models \cite{Totani:1995rg,Malaney:1996ar,Hartmann:1997qe,Kaplinghat:1999xi,Fukugita:2002qw,Ando:2005,Horiuchi:2008jz,Lunardini:2009ya,Galais:2009wi,Nakazato:2015rya,Priya:2017bmm,Horiuchi:2017qja,Barranco:2017lug,DeGouvea:2020ang,Horiuchi:2020jnc,Kresse:2020nto,Tabrizi:2020vmo,Ivanez-Ballesteros:2022szu,Ashida:2023heb,Martinez-Mirave:2024zck}
that were used in the most recent SK data analysis \cite{Super-Kamiokande:2025sxh} \footnote{We restricted to those models in the SK analysis for which 
the energy spectrum 
could  be reliably reconstructed,
by cross-correlating the model spectra shown in Ref. \cite{Super-Kamiokande:2025sxh} with the spectra in the original model references. }.
The same notation as in the SK analysis will be used;
see Supplemental Material for details.  We 
note that the models are very diverse: they use different degrees of approximations and include different effects. Some of these 
effects 
are expected to break the exponential approximation: 
for example, neutrino oscillations could lead to deviations from the alpha spectrum, and the assumption of identical emitters is violated when including powerful sources like black hole-forming collapses (failed \sne) \cite{Lunardini:2009ya,Nakazato:2015rya,Ashida:2023heb}, magnetorotational \sne\ \cite{Martinez-Mirave:2024zck}, fallback accretion \cite{Nakazato:2024gem}, etc., which typically have hotter \n\ spectra. 
Some models (see Ref. \cite{Ivanez-Ballesteros:2022szu}) include physics beyond the Standard Model for which the exponential description is generally not expected to hold. Still, in all these cases the exponential spectrum turns out to be a good approximation, as will be seen below.

For each model, tabulated spectra\footnote{Most tables were obtained by digitizing published figures.
The error induced by the digitization should be minor and not affect our conclusions.}
  are fit to 
Eq. (\ref{eq:exp}) using least-squares minimization, so to obtain the best fit values of the parameters, ${\mathcal{E}_0}$ and $\phi_{20}$. We only included tabulated points with $E\geq 10$ MeV, to account for the energy window of interest, and perform the fit in logarithmic space due to the large range of values of $\phi(E)$. For all models, the quality of the fit is very good, with the coefficient of determination, $R^2$, being very close to 1: $ 1-R^2 \leq 0.01$. 
We find that  $R^2$ is smaller for those models where some breaking of the exponential approximation is expected.  
 In Fig. \ref{fig:dsnb_expofit_five_panels}, the tabulated spectra and the corresponding best-fit exponential curves are shown; the good agreement 
 is evident.

The exponential approximation allows every model to be mapped onto a point in the two-dimensional parameter space $(\phi_{20},\mathcal{E}_0)$, as shown in Fig.~\ref{fig:dsnb_f3_integrated_flux_vs_e0}(b) (see also Table \ref{tab:dsnb_f3_f4_summary} in the Supplemental Material).
Interestingly, we find that, while the models originate from a wide variety of physical assumptions, their spectral parameter, $\mathcal{E}_0$, occupies a relatively restricted interval: we get $3.35 \leq \mathcal{E}_0/\mathrm{ MeV} \leq 8.0$, with 21 out of 24 models lying in the interval $4.7 \leq \mathcal{E}_0/\mathrm {MeV} \leq 6.0$. 

\begin{figure*}[htbp]
\centering
\includegraphics[width=0.95 \textwidth]{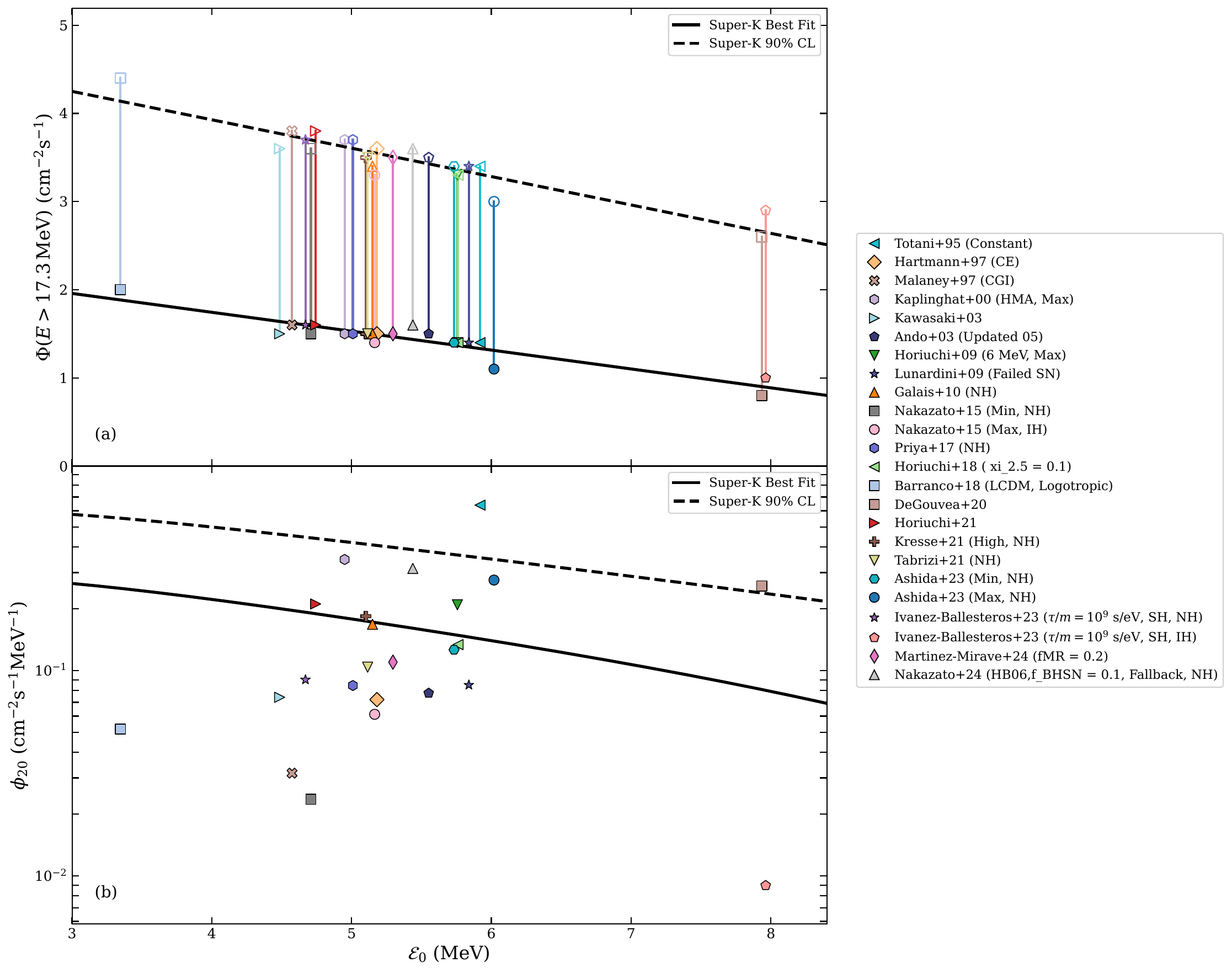} 
\caption{{\bf (a):} Models for the $\barnue$ component of the
\df, in the plane of integrated flux above the Super-Kamiokande  
energy threshold, $E_{th}=17.3$ MeV, 
versus the exponential scale parameter, $\mathcal{E}_0$.
For each model, the filled and open markers indicate the Super-Kamiokande best-fit flux and 90\% C.L. upper limit, respectively (NN-based analysis in Ref. \cite{Super-Kamiokande:2025sxh}).
The solid (dashed) black line shows the linear fit to the Super-Kamiokande best-fit (90\% C.L. upper limits) values as a function of $\mathcal{E}_0$. 
{\bf (b):} the same models in the plane of their best-fitting exponential spectrum parameters, $(\phi_{20},\mathcal{E}_0)$ (see eq. (\ref{eq:exp})). The solid (dashed) black curve shows our derived Super-Kamiokande  best-fit (90\% upper bound) for $\phi_{20}$ as a function of $\mathcal{E}_0$.
}
\label{fig:dsnb_f3_integrated_flux_vs_e0}
\end{figure*}
The utility of the $(\phi_{20},\mathcal{E}_0)$ framework becomes particularly evident when interpreting experimental results. For each of the models considered here, the recent SK analysis \cite{Super-Kamiokande:2025sxh} reported best-fit values and 90\% CL upper limits for the integrated $\barnue$ component of the \df, $\Phi$, above the neutrino energy threshold of $E_{th}=17.3$ MeV. These results were obtained by performing one-parameter fits of each model to the data, where the flux normalization was the fitting parameter and the  spectrum was kept fixed. 
We consider the best fits and upper bounds from the Neural Network (NN)-based analysis, see Table F3 in Ref. \cite{Super-Kamiokande:2025sxh}. Results are nearly identical for the analysis using the Boosted Decision Tree (BDT) method; they are given in the Supplemental Material.
Figure~\ref{fig:dsnb_f3_integrated_flux_vs_e0}(a) shows how the upper bounds and best fits  for the individual models correlate with $\mathcal{E}_0$ (plotted on the horizontal axis).
 We observe a clear trend, where both upper limits and best fit values decrease with $\mathcal{E}_0$. 
This behavior agrees with what is expected phenomenologically, because a constant number of events above the threshold can be reproduced by either increasing $\mathcal{E}_0$ or 
lowering the flux normalization. 
The trend is approximately linear\footnote{We also considered a  function corresponding to a fixed number of events in SK. The quality of the fit was comparable to the linear fit, which was preferred for its simplicity. }; from a least squares fit we find that it is well reproduced by $\Phi_{\rm best}= 2.601 - 0.214\,\mathcal{E}_0$ and $\Phi_{90\%}= 5.215   -0.322 \,\mathcal{E}_0$  for the best fits and upper bounds respectively.

Having established that the exponential form reproduces models well, and is effective to interpret the SK results, we can go a step further and express the SK results as best fits and bounds on $\phi_{20}$, for a given (fixed) value of $\mathcal{E}_0$. This is done by solving the equation $\Phi_\mathrm{ best}=\int_{E_{th}}^\infty \phi(E)dE$ (and similar for $\Phi_{90\%}$) for $\phi_{20}$.  
 Fig.~\ref{fig:dsnb_f3_integrated_flux_vs_e0}(b) shows the result of this exercise. Best fits and bounds are shown as continuous functions of $\mathcal{E}_0$, and compared with the predictions of the individual models.
This figure offers an immediate and compact view of how the theoretical landscape compares with the present experimental situation. We see that predictions are largely consistent with the data, with only two models being in (mild) tension.

The simplicity of the exponential description suggests a broad role for it in future \df\ studies. Rather than reporting only numerical spectra, theoretical calculations could routinely quote the values of $(\phi_{20},\mathcal{E}_0)$ that best approximate them, thus facilitating comparisons between models and with experimental constraints. Some authors have already adopted this approach \cite{Kresse:2020nto}; we hope that it will become standard practice.

Similarly, future experimental searches could use the exponential parameterization as a standard framework.  They could go beyond the current one-parameter fits and perform \emph{two}-parameter fits directly in the $(\phi_{20},\mathcal{E}_0)$ plane. Reporting likelihoods or regions of confidence in this plane would make the results immediately applicable to any model admitting an exponential description, enable direct comparisons between the results of different searches, and facilitate global statistical analyses. Such consistency will become increasingly useful when the SK results are complemented by JUNO and Hyper-Kamiokande. Moreover, since $\mathcal{E}_0$ can be approximately related to the \n\ spectrum of an individual supernova \cite{Lunardini:2006pd}, \df\ constraints could be compared with fits to the $\barnue$ signal from SN1987A (see, e.g., \cite{Lunardini:2005jf}).

The usefulness of an effective description is illustrated by the experience of TeV-PeV \n\ astronomy at IceCube \cite{IceCube:2013cdw,IceCube:2013low}. There, a phenomenological power-law spectrum has supported 
the field from discovery to precision measurements (see, e.g., \cite{Kappes:2025jag}). Only recently deviations from a power law have begun to emerge \cite{IceCube:2025tgp}. We envision an analogous role for the exponential parameterization of the \df. Its simplicity could also facilitate applications beyond \df\ astrophysics, including particle phenomenology beyond the Standard Model and the development of new \n\ detection concepts.

The exponential form should  be regarded as an \emph{effective} description of the current observational window. At $E\lesssim 10$ MeV, a suppression is expected as the energy approaches the \df\ peak at $E\sim4$–$6$ MeV \cite{Lunardini:2006pd}.  At higher energies, ${\mathcal O}(10^{-2})$ corrections 
are predicted due to flavor oscillations inside the Earth (see, e.g. \cite{Ando:2002zj}), and potentially significant
deviations could arise from black-hole-forming collapses, exotic flavor-conversion effects, or other physics beyond the Standard Model. As \df\ observations enter a precision phase, with lower thresholds, better-controlled backgrounds, and higher statistics, the exponential spectrum can serve as a natural null hypothesis for searches of such features.

The present analysis could be extended to other \n\ flavors, particularly the $\nue$ component of the \df\ that will be probed at DUNE, for which we expect similar results. A larger compilation of $(\phi_{20},\mathcal{E}_0)$ values would also help characterize more completely the parameter space favored by theory.

In summary, we have shown that a broad collection of modern \df\ models is accurately represented, within the observable energy window, by a simple exponential spectrum. Above realistic detection thresholds, the \df\ is therefore effectively a two-parameter phenomenon, characterized by its normalization $\phi_{20}$ and spectral scale $\mathcal{E}_0$. This framework provides a common language for comparing theoretical models and experimental measurements and for broader phenomenological applications. We advocate its adoption in future \df\ predictions and searches, while future high-statistics analyses test for departures from it.

\begin{acknowledgments}

We are grateful to Hiroyuki Sekiya, Saki Fujita, and Masayuki Harada for valuable discussions and for kindly sharing information about the Super-Kamiokande analysis. We also thank John Beacom and Mark Vagins for fruitful discussions.
This work was supported by the NSF awards  No.~2309973 and 2609687. 

\noindent{}
\emph{Artificial Intelligence (AI) use disclosure.} Part of the numerical analysis described here was assisted by ChatGPT 5.6. The analysis was closely directed by the authors, and the results have been verified by the authors through independent calculations and cross-checks. ChatGPT 5.6 also produced the Python codes that were used to prepare the figures, and  assisted with the initial drafting of the manuscript. The authors reviewed and edited both these codes and the text as needed, and take full responsibility for them.

\end{acknowledgments}

\bibliography{refs}

%%%%%%%%%%%%%%%%%%%%%%%%%%%%%%%%%%%%%%%%%%%%%%%%%%

% Supplemental Material

%%%%%%%%%%%%%%%%%%%%%%%%%%%%%%%%%%%%%%%%%%%%%%%%%%

\clearpage

\onecolumngrid

\begin{center}

{\large\bf SUPPLEMENTAL MATERIAL}

\end{center}

\vspace{0.5cm}

%\section*{Supplemental Material for ``A Two-Parameter Framework for the Diffuse Supernova Neutrino Background''}

This Supplemental Material contains additional figures and tables supporting the analysis presented in the main text. In particular, we present a table where, for each theoretical model, we report the best fitting exponential spectrum parameters, $\phi_{20}$ and $\mathcal{E}_0$, and the results of the most recent Super-Kamiokande analysis (for both the NN-based and the BDT-based methods).  An additional figure is provided as well, illustrating the results of our work when the BDT-based Super-Kamiokande best fits and upper limits are used.

\begin{table}[htbp]
\centering
\small
\setlength{\tabcolsep}{4pt}
\caption{Exponential fit parameters for the 24 representative \df\ models analyzed in this work ($\barnue$ component only). To identify the models, we use the same notation as in Ref. \cite{Super-Kamiokande:2025sxh}, to which we refer for details; here it suffices to say that NH and IH refer to flavor oscillations with the normal and inverted mass hierarchy, respectively. For each model, we also give the corresponding Super-Kamiokande NN- and BDT-based best-fits and 90\% C.L. upper limits (UL$_{90}$) for the integrated $\barnue$ flux above the threshold $E_{th}=17.3$ MeV. The NN-based and BDT-based results are taken from tables F3 and F4 in Ref. \cite{Super-Kamiokande:2025sxh}, respectively.
This table provides the mapping between published \df\ models and the two-parameter phenomenological framework introduced in the main text. }
\label{tab:dsnb_f3_f4_summary}
\begin{tabular}{lcccccc}
\hline
 Model & $\mathcal{E}_0$  & $\Phi_{20}$ & NN Best & NN UL$_{90}$ & BDT Best  & BDT UL$_{90}$ \\
 & (MeV) & ($\mathrm{ cm^{-2} s^{-1}}$ & ($\mathrm{ cm^{-2} s^{-1}}$) & ($\mathrm{ cm^{-2} s^{-1}}$) & ($\mathrm{ cm^{-2} s^{-1}}$) & ($\mathrm{ cm^{-2} s^{-1}}$) \\
  &   & $\mathrm{  MeV^{-1}}$) &  &  &  &  \\
\hline
Ando+03 (Updated 05) \cite{Ando:2005}  & 5.552 & 0.07762 & 1.5 & 3.5 & 1.2 & 3.5 \\
Ashida+23 (Max, NH) \cite{Ashida:2023heb} & 6.019 & 0.2757 & 1.1 & 3.0 & 0.9 & 3.0 \\
Ashida+23 (Min, NH)  \cite{Ashida:2023heb} & 5.733 & 0.1263 & 1.4 & 3.4 & 1.2 & 3.4 \\
Barranco+18 (LCDM, Logotropic) \cite{Barranco:2017lug} & 3.346 & 0.05189 & 2.0 & 4.4 & 1.7 & 4.2 \\
DeGouvea+20\footnote{This publication includes some predictions with physics beyond the Standard Model, including pseudo-Dirac neutrinos and neutrino decay.} \cite{DeGouvea:2020ang} & 7.935 & 0.2577 & 0.8 & 2.6 & 0.7 & 2.7 \\
Galais+10 (NH) \cite{Galais:2009wi} & 5.149 & 0.1679 & 1.5 & 3.4 & 1.3 & 3.4 \\
Hartmann+97 (CE) \cite{Hartmann:1997qe}  & 5.181 & 0.07217 & 1.5 & 3.6 & 1.3 & 3.5 \\
Horiuchi+09 (6 MeV, Max) \cite{Horiuchi:2008jz} & 5.757 & 0.209 & 1.4 & 3.3 & 1.2 & 3.3 \\
Horiuchi+18 ( $\xi_{2.5}$ = 0.1) \cite{Horiuchi:2017qja} & 5.761 & 0.1337 & 1.4 & 3.3 & 1.2 & 3.3 \\
Horiuchi+21 \cite{Horiuchi:2020jnc} & 4.743 & 0.2109 & 1.6 & 3.8 & 1.4 & 3.7 \\
Ivanez-Ballesteros+23 ($\tau/m= 10^9~ \mathrm{ s/eV}$, SH, IH)\footnote{This model involves physics Beyond the Standard Model; specifically, neutrino decay. } \cite{Ivanez-Ballesteros:2022szu} & 7.962 & 0.008985 & 1.0 & 2.9 & 1.0 & 2.9 \\
Ivanez-Ballesteros+23 ($\tau/m= 10^9~ \mathrm{ s/eV}$, SH, NH) \cite{Ivanez-Ballesteros:2022szu} & 4.670 & 0.09021 & 1.6 & 3.7 & 1.4 & 3.7 \\
Kaplinghat+00 (HMA, Max) \cite{Kaplinghat:1999xi} & 4.950 & 0.3473 & 1.5 & 3.7 & 1.3 & 3.6 \\
Kawasaki+03 \cite{Fukugita:2002qw} & 4.485 & 0.07406 & 1.5 & 3.6 & 1.3 & 3.6 \\
Kresse+21 (High, NH) \cite{Kresse:2020nto}  & 5.103 & 0.1838 & 1.5 & 3.5 & 1.3 & 3.5 \\
Lunardini09 (Failed SN)\footnote{Model for LS equation of state and $\bar p=0.68$, bottom left panel of Fig. 2 in Ref. \cite{Lunardini:2009ya}.} \cite{Lunardini:2009ya} & 5.839 & 0.08505 & 1.4 & 3.4 & 1.2 & 3.5 \\
Malaney97 (CGI) \cite{Malaney:1996ar} & 4.574 & 0.03164 & 1.6 & 3.8 & 1.6 & 3.7 \\
Martinez-Mirave+24 ($f_\mathrm{MR} = 0.2$) \cite{Martinez-Mirave:2024zck} & 5.296 & 0.1099 & 1.5 & 3.5 & 1.3 & 3.4 \\
Nakazato+15 (Max, IH) \cite{Nakazato:2015rya} & 5.165 & 0.06121 & 1.4 & 3.3 & 1.2 & 3.3 \\
Nakazato+15 (Min, NH) \cite{Nakazato:2015rya} & 4.709 & 0.0236 & 1.5 & 3.6 & 1.3 & 3.6 \\
Nakazato+24 (HB06, $f_\mathrm{ BHSN}$ = 0.1, Fallback, NH) \cite{Nakazato:2024gem} & 5.438 & 0.3142 & 1.6 & 3.6 & 1.4 & 3.5 \\
Priya+17 (NH) \cite{Priya:2017bmm} & 5.009 & 0.08454 & 1.5 & 3.7 & 1.3 & 3.6 \\
Tabrizi+21 (NH) \cite{Tabrizi:2020vmo} & 5.115 & 0.1038 & 1.5 & 3.5 & 1.2 & 3.5 \\
Totani+95 (Constant) \cite{Totani:1995rg} & 5.919 & 0.6385 & 1.4 & 3.4 & 1.2 & 3.4 \\
\\
\hline
\end{tabular}
\end{table}

\begin{figure*}[htbp]
\centering
\includegraphics[width=0.95 \textwidth]{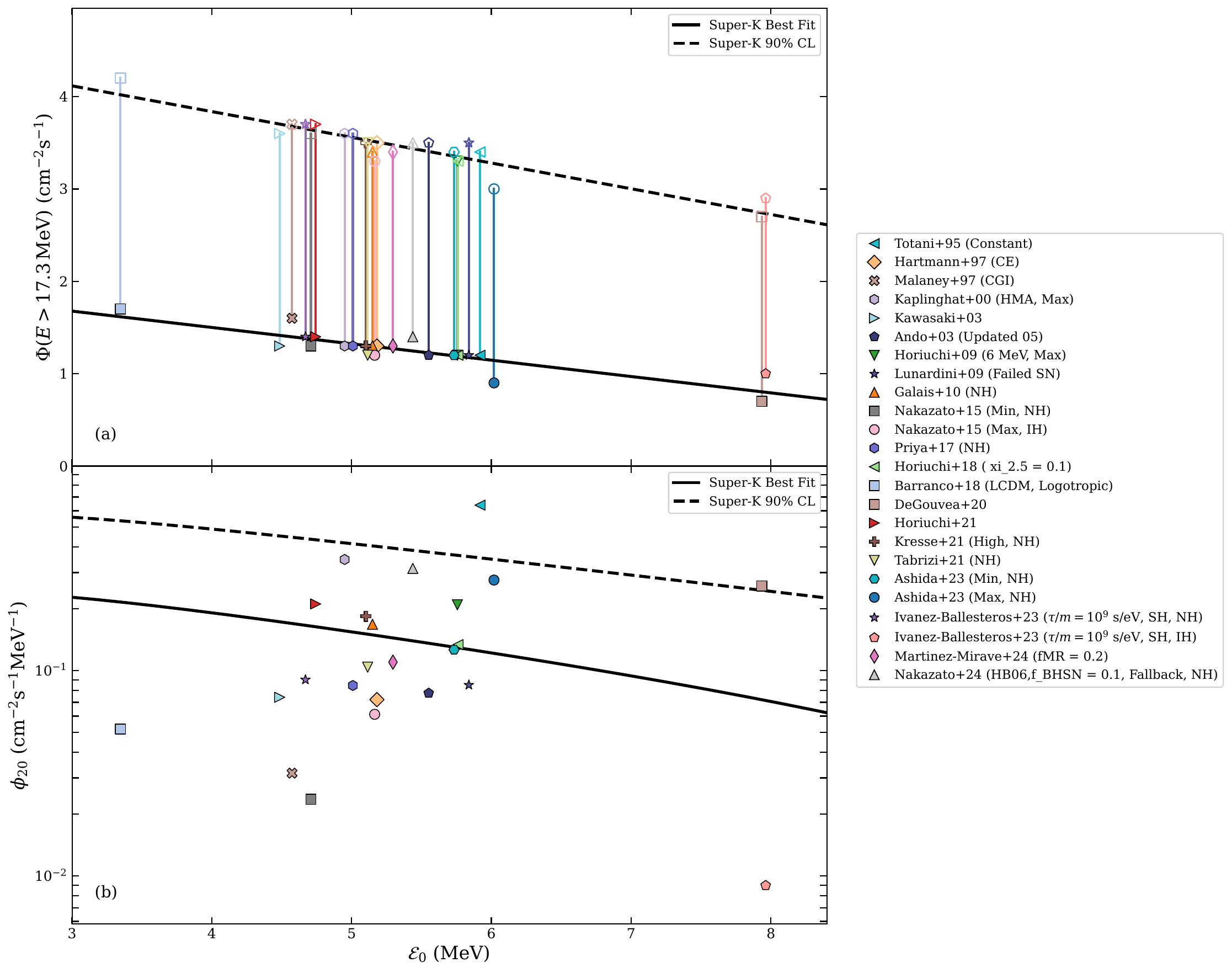} 
\caption{{\bf (a):} Models for the $\barnue$ component of the
\df, in the plane of integrated flux above the Super-Kamiokande  
energy threshold, $E_{th}=17.3$ MeV, 
versus the exponential scale parameter, $\mathcal{E}_0$.
For each model, the filled and open markers indicate the Super-Kamiokande best-fit flux and 90\% C.L. upper limit, respectively (BDT-based analysis in Ref. \cite{Super-Kamiokande:2025sxh}).
The solid and dashed lines show the linear fit to the Super-Kamiokande best-fit and 90\% C.L. upper limits, respectively, as a function of $\mathcal{E}_0$. 
Their equations are: $\Phi_{\rm best}=  2.209-0.177 \mathcal{E}_0$ and $\Phi_{90\%}=4.951  -0.278\,\mathcal{E}_0$.
{\bf (b):} the same models in the plane of their best-fitting exponential spectrum parameters, $(\phi_{20},\mathcal{E}_0)$ (see eq. (\ref{eq:exp})). The solid (dashed) black curve shows our derived Super-Kamiokande  best-fit (90\% upper bound) for $\phi_{20}$ as a function of $\mathcal{E}_0$.
}
\label{fig:dsnb_f4_integrated_flux_vs_e0}
\end{figure*}

\end{document}